\documentclass[reprint,superscriptaddress, amsmath,amssymb,aps, prl]{revtex4-2}

\usepackage{graphicx}% Include figure files
\usepackage{dcolumn}% Align table columns on decimal point
\usepackage{bm}% bold math
\usepackage{amsmath,mathtools}
\usepackage{hyperref}% add hypertext capabilities
\begin{document}

\title{\textbf{Alter-excitons in Altermagnets: Symmetry, Splitting and Dispersion} 
}% 

\author{Yuanfan Xiong}
\affiliation{Hefei National Laboratory, University of Science and Technology of China, Hefei, Anhui 230088, China.}
\affiliation{School of emerging technology, University of Science and Technology of China, Hefei, Anhui 230026, China.}
\affiliation{Hefei National Research Center for Physical Sciences at the Microscale, University of Science and Technology of China, Hefei, Anhui 230026, China.}

\author{Liangyu Wang}
\affiliation{Hefei National Research Center for Physical Sciences at the Microscale, University of Science and Technology of China, Hefei, Anhui 230026, China.}

\author{Xuan Wu}
\affiliation{Hefei National Laboratory, University of Science and Technology of China, Hefei, Anhui 230088, China.}
\affiliation{School of emerging technology, University of Science and Technology of China, Hefei, Anhui 230026, China.}
\affiliation{Hefei National Research Center for Physical Sciences at the Microscale, University of Science and Technology of China, Hefei, Anhui 230026, China.}

\author{Wei Hu}%
\email{Contact author: whuustc@ustc.edu.cn}
\affiliation{Hefei National Laboratory, University of Science and Technology of China, Hefei, Anhui 230088, China.}
\affiliation{School of emerging technology, University of Science and Technology of China, Hefei, Anhui 230026, China.}
\affiliation{Hefei National Research Center for Physical Sciences at the Microscale, University of Science and Technology of China, Hefei, Anhui 230026, China.}

\author{Jinlong Yang}
\email{Contact author: jlyang@ustc.edu.cn}
\affiliation{State Key Laboratory of Precision and Intelligent Chemistry, and Department of Chemical Physics, University of Science and Technology of China, Hefei, Anhui 230026, China.}

\date{\today}% It is always \today, today,
             %  but any date may be explicitly specified

\begin{abstract}
Magnetism provides a fertile platform for discovering emergent quantum phenomena governed by symmetry. Recent developments of spin-space-group (SSG) theory have revealed altermagnets as a distinct class of nonrelativistic collinear magnets with unconventional spin-dependent electronic structures. Here, we demonstrate that altermagnetic symmetry imposes fundamentally different constraints on electronic and excitonic excitations. Although the quasiparticle and two-body Hamiltonians possess identical SSG symmetries, we find that symmetry-protected degeneracies present in the electronic band structure can be intrinsically lifted in the exciton spectrum at specific momenta. This splitting arises solely from the many-body nature of excitons and does not require spin-orbit coupling or noncollinear magnetic order. Combining symmetry analysis with many-body perturbation theory, we predict a novel class of spin-split excitonic quasiparticles, termed alter-excitons, and reveal their characteristic dispersions in altermagnetic systems. Our results establish excitons as a sensitive probe of hidden spin-space symmetries and uncover a new route toward symmetry-engineered many-body phenomena in collinear magnets.
\end{abstract}
\maketitle

%\tableofcontents

Magnetism is one of the central themes of condensed-matter physics. Conventionally, collinearly ordered magnets are classified as ferromagnets, which possess a finite macroscopic magnetization, or antiferromagnets, in which oppositely polarized sublattices compensate to yield a vanishing net moment. The recently developed spin-space-group (SSG) framework~\cite{chen_enumeration_2024,jiang_enumeration_2024,liu_spin-group_2022,xiao_spin_2024,song_constructions_2025}, which treat spin rotations and real-space operations separately, has revealed a distinct third class of nonrelativistic collinear magnets—altermagnets. Altermagnets combine compensated antiparallel magnetic order with momentum-dependent spin splitting, thereby exhibiting characteristics conventionally associated with both ferromagnets and antiferromagnets~\cite{gonzalez-hernandez_efficient_2021,smejkal_beyond_2022,smejkal_emerging_2022,bai_altermagnetism_2024,chen_unconventional_2025}. In altermagnets, the compensated antiparallel magnetic order does not enforce spin degeneracy throughout the entire Brillouin zone; instead, it introduces a $\mathbf{k}$-dependent spin splitting that is intimately linked to the crystal rotation $A$ connecting the opposite-spin sublattices. Quasi-Kramers spin degeneracy survives only at momenta where the little group $\boldsymbol{\mathrm{L_k}}$ contains real-space transformations between spin-up and spin-down channels~\cite{smejkal_beyond_2022,smejkal_emerging_2022,chen_unconventional_2025}. The unconventional electronic structure of altermagnets provides a unique platform for exploring many quantum phenomena, including the anomalous Hall effect~\cite{smejkal_crystal_2020,ghimire_large_2018,feng_anomalous_2022,gonzalez_betancourt_spontaneous_2023}, spin-splitter torque~\cite{song_altermagnets_2025,karube_observation_2022,bai_observation_2022}, and superconductivity~\cite{smejkal_emerging_2022,sun_andreev_2023,ouassou_dc_2023,zhang_finite-momentum_2024,chakraborty_zero-field_2024}. The intriguing research landscape of altermagnets makes this exotic magnetic phase a compelling topic in condensed matter physics.

Considerable effort has therefore been devoted to identifying altermagnetic materials with tailored symmetries and characterizing their single-particle electronic structures, resulting in theoretical predictions and experimental evidence for candidate systems including $\mathrm{RuO_2}$, $\mathrm{MnTe}$, and $\mathrm{V_2Se_2O}$~\cite{song_altermagnets_2025,fender_altermagnetism_2025,bai_altermagnetism_2024,lee_broken_2024,yang_ultrafast_2026}. By contrast, the consequences of altermagnetic symmetry for interacting electron–hole excitations remain largely unexplored. Scattered efforts have addressed bright excitons at zero momentum~\cite{sun_optically_2025,cao_symmetry_2025}. Excitons are composite bosons formed by electron–hole pairs bound via the Coulomb interaction. As elementary excitations in matter, they play a crucial role in determining electronic and optical properties, with their internal symmetries dictating selection rules, coherent dynamics, and many-body interactions. Importantly, an exciton inherits neither its symmetry nor its dispersion trivially from the underlying single-particle bands. Extending altermagnetic symmetry concepts from single-particle states to excitonic quasiparticles can consequently reveal phenomena inaccessible at the independent-particle level, including symmetry-modified exciton degeneracies, spin- and polarization-selective optical transitions, and spin-dependent exciton propagation. Excitons in altermagnets thus provide a natural many-body arena in which to probe how altermagnetic spin splitting is reshaped by electron–hole correlations, while also opening opportunities for spin-selective optoelectronic and quantum functionalities.

In this letter, we demonstrate that, although the single-particle and two-body Hamiltonians of an altermagnetic system share the same spin-space-group symmetry, degeneracies protected in the electronic band structure at specific $\mathbf{k}$ points need not persist in the excitonic spectrum. Our many-body perturbation theory results confirm this symmetry-based prediction and reveal a characteristic momentum-dependent splitting and alternating structure of the resulting alter-excitons. We further identify the electron–hole exchange interaction as the microscopic origin of the excitonic splitting. We also investigate the optical effects and the exciton-wave-packet dynamics by the GW+BSE method and the TD-aGW method, establishing alter-excitons as a new class of many-body excitations with optical and transport phenomena beyond those encoded in the underlying single-particle bands.
\begin{figure}[!htb]
    \centering
    \includegraphics[width=1.0\linewidth]{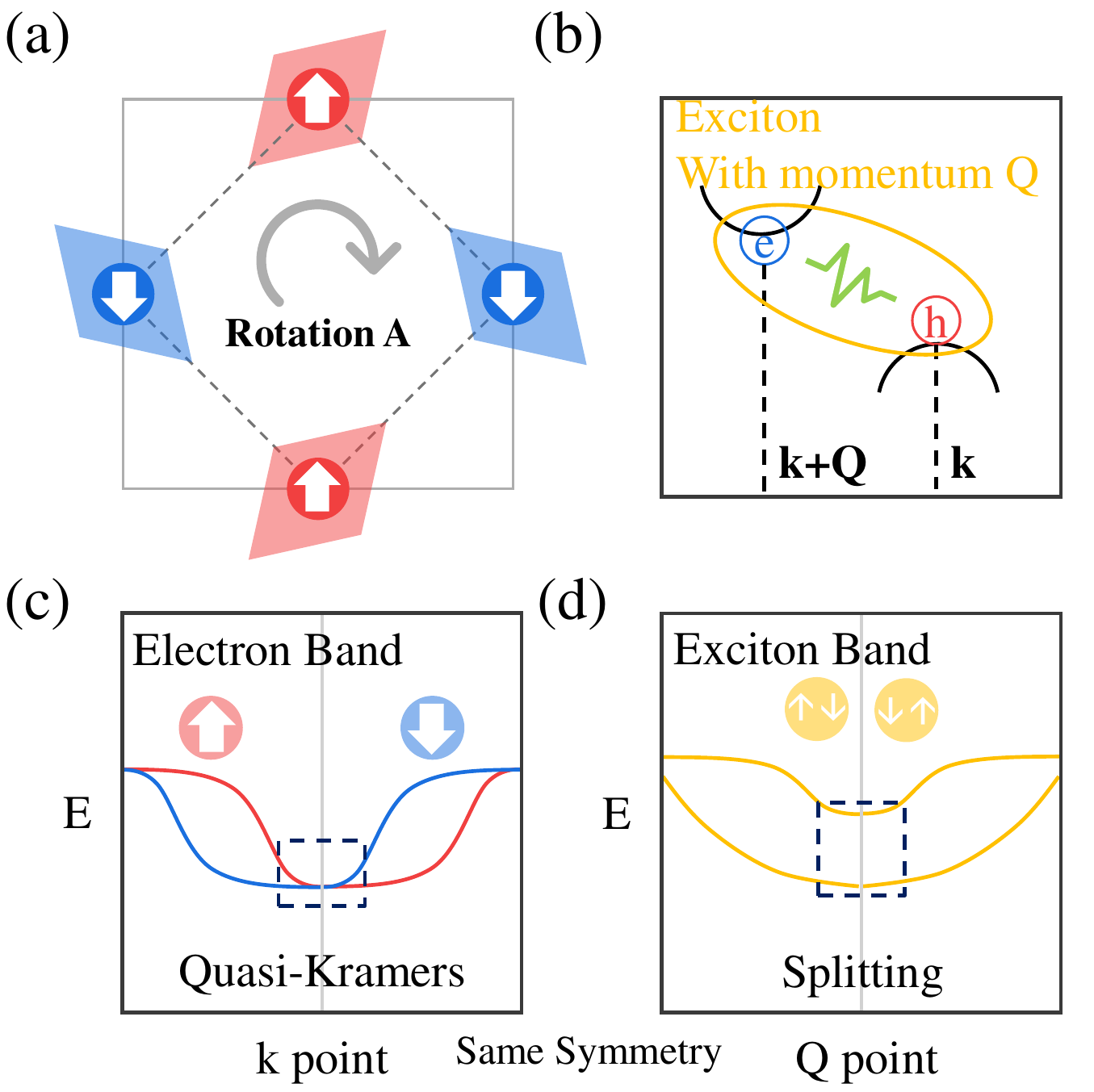}
    \caption{Schematic diagrams for altermagnetism, exciton and band structure. (a) The symmetry operation $\{\mathcal{T}||A\}$ or $\{U_x(\pi)||A\}$ connects spin-up and spin-down sublattice, which induces altermagnetism. (b) Schematic of an exciton with center-of-mass momentum $\mathbf{Q}$ (c) Electronic band structure of altermagnets. Quasi-Kramers degeneracies emerge at some high-symmetry momenta. (d) Excitonic band structure of altermagnets. Although exciton holds the same little group as electron at the same momenta, a corresponding degeneracy in the excitonic band is not enforced.}
    \label{fig:fig1}
\end{figure}

We begin by reviewing the symmetry origin of the degeneracy in altermagnets. The symmetry of altermagnets is described by the type-$\mathrm{III}$ collinear spin space group $G_{\mathrm{SSG}}^\mathrm{III}=G_{\mathrm{NSSG}}\times G_{\mathrm{SO}}$, where the spin-only part is $G_{SO}=Z_2^K \ltimes\mathrm{SO}(2)$ and $G_{\mathrm{SSG}}$ is indeed the quotient group with respect to the translation subgroup $\mathbb{T}$. For altermagnets, $G_{\mathrm{NSSG}}$ can be decomposited by the cosets of spin-sublattice group $G_\uparrow$: $G_{\mathrm{NSSG}}=\{E||G_\uparrow\}\cup\{\mathcal{T}||AG_\uparrow\}$. The degeneracy of electronic bands at specific $\mathbf{k}$-points is exhibited by the $\mathcal{T}A$ operations connecting spin sublattices. For collinear SSG, non-trivial spin space operations include $\mathcal{T}, U_x(\pi)$ and $U_z(\theta)$ and their combinations. In electron systems they have representation matrix as:
\begin{equation}
    D(\mathcal{T})=\left(\begin{matrix}0&1\\-1&0\end{matrix}\right)K,
\end{equation}
\begin{equation}
    D(U_z(\theta))=\left(\begin{matrix}e^{-i\theta/2}&0\\0&e^{i\theta/2}\end{matrix}\right),
\end{equation}
where complex conjugate operator $K$ enters $\mathcal{T}$. Two mechanisms give rise to band degeneracy at specific $\mathbf{k}$ points in the electronic structure. First, based on the Dimmock-Wheeler character sum rule, the summation of character $\chi([\mathcal{T}AU_z(\theta)]^2)$ gives zero (see Ref.~\cite{chen_enumeration_2024}), so the symmetry operation $\mathcal{T}AU_z(\theta)$ pairs two conjugated one-dimensional irreducible representations into a two-dimensional irreducible representation. This contributes to an extra degeneracy and is the origin of the quasi-Kramers degeneracy in altermagnets at some high-symmetry points. Second, the combined action of $U_x(\pi)$ and $U_z(\phi)$ contributes to an emergence of the infinite group $D_\infty$, whose two-dimensional irreducible representations directly enforce the two-fold degeneracy. Thus, spin degeneracy appears at $\mathbf{k}$ points whose little group contains either the symmetry operation $\{U_x(\pi)||A\}$ or $\{\mathcal{T}||A\}$, while spin-splitting occurs at the other $\mathbf{k}$ points.

Then we performed the symmetry analysis for excitons. When dealing with the excitons in altermagnets, the spin configurations become complex. We use the Bethe-Salpeter equation (BSE) to describe the two-body bound states~\cite{strinati_application_1988,rohlfing_electron-hole_2000}. The Bethe-Salpeter equation $L=L_0+L_0KL$ is invariant under the symmetry operations of the system. Considering our crystal system, we can simplify the equation into an eigenvalue problem by adopting the GW approximation. The two-body Hamiltonian with spin configurations is~\cite{rohlfing_electron-hole_2000}

\begin{equation}
\begin{aligned}
    H&=\left(\begin{matrix}
        H_{|M_S|=0}&\mathbf{0}\\
        \mathbf{0}&H_{|M_S|=1}
    \end{matrix}\right)\\&=\left(\begin{matrix}
        H^0_{\uparrow\uparrow,\uparrow\uparrow}&v_{\uparrow\uparrow,\downarrow\downarrow}&0&0\\
        v_{\downarrow\downarrow,\uparrow\uparrow}&H^0_{\downarrow\downarrow,\downarrow\downarrow}&0&0\\
        0&0&H^1_{\uparrow\downarrow,\uparrow\downarrow}&0\\
        0&0&0&H^1_{\downarrow\uparrow,\downarrow\uparrow}
    \end{matrix}\right), 
\end{aligned}
\end{equation}
where $H^0=D+W+v$ and $H^1=D+W$. $D, W$, and $v$ are the single particle energy difference, direct kernel, and exchange kernel, respectively. The exciton states are expanded on the products of the Bloch states: 
\begin{equation}
    \Psi_\mathrm{\boldsymbol{Q}}^S(\mathrm{\boldsymbol{r}_e},\mathrm{\boldsymbol{r}_h})=\sum_{\substack{n_1n_2\mathrm{\boldsymbol{k}} \\ \sigma_1\sigma_2}}{A_{\substack{n_1n_2\mathrm{\boldsymbol{k}}\\\sigma_1\sigma_2}}^S\psi^*_{n_1\mathrm{\boldsymbol{k-Q}},\sigma_1}(\mathrm{\boldsymbol{r}_h})\psi_{n_2\mathrm{\boldsymbol{k}},\sigma_2}(\mathrm{\boldsymbol{r}_e})}.
\end{equation}
The basis is $\psi^*_{n_1\mathrm{\boldsymbol{k-Q}},\sigma_1}(\mathrm{\boldsymbol{r}_h})\psi_{n_2\mathrm{\boldsymbol{k}},\sigma_2}(\mathrm{\boldsymbol{r}_e})$, where the spin indices $\sigma_1$ and $\sigma_2$ are spin indices of the corresponding electron states. Thus, the spin indices for holes are indeed $\sigma_h=-\sigma_1$. So the total $M_s$ for exciton is $M_s=\sigma_e+\sigma_h=\sigma_2-\sigma_1$. Clearly, the Hamiltonian can be divided into two parts according to the spin components of the basis. The first block $H_{|M_S|=0}$ contains the $\sigma_1=\sigma_2$ bases and the second block $H_{|M_S|=1}$ contains the $\sigma_1=-\sigma_2$ bases. The $H_{|M_S|=1}$ block yields two spin-triplet states of the spin-flip type, with $S=1$ and $M_s=\pm 1$. To examine the existence of extra degeneracy, we apply the operation $\mathcal{T}AU_z(\theta)$ to the basis $[\psi^*_\uparrow\psi_\downarrow,\psi^*_\downarrow\psi_\uparrow]$ (here the spin indices indicate the spin of the corresponding electron states). We have
\begin{equation}
    [\mathcal{T}U_z(\theta)]^2\left(\begin{array}{c}
         \psi^*_\uparrow\psi_\downarrow\\
         \psi^*_\downarrow\psi_\uparrow
    \end{array}\right)=\left(\begin{array}{cc}
         e^{-2i\theta}\psi^*_\uparrow\psi_\downarrow\\
         e^{2i\theta}\psi^*_\downarrow\psi_\uparrow
    \end{array}\right),
\end{equation}
which is the same as previous research on magnons in altermagnets\cite{chen_unconventional_2025}. Indeed, this type of excitation can be identified as a magnon excitation, as demonstrated in Ref.~\cite{marsili_spinorial_2021}. Previous studies on unconventional magnons within the collinear spin-space-group framework have reported this findings~\cite{chen_unconventional_2025}. In this work, we mainly focus on the excitonic block $H_{|M_S|=0}$ below.

\begin{figure}[!htb]
    \centering
    \includegraphics[width=1.0\linewidth]{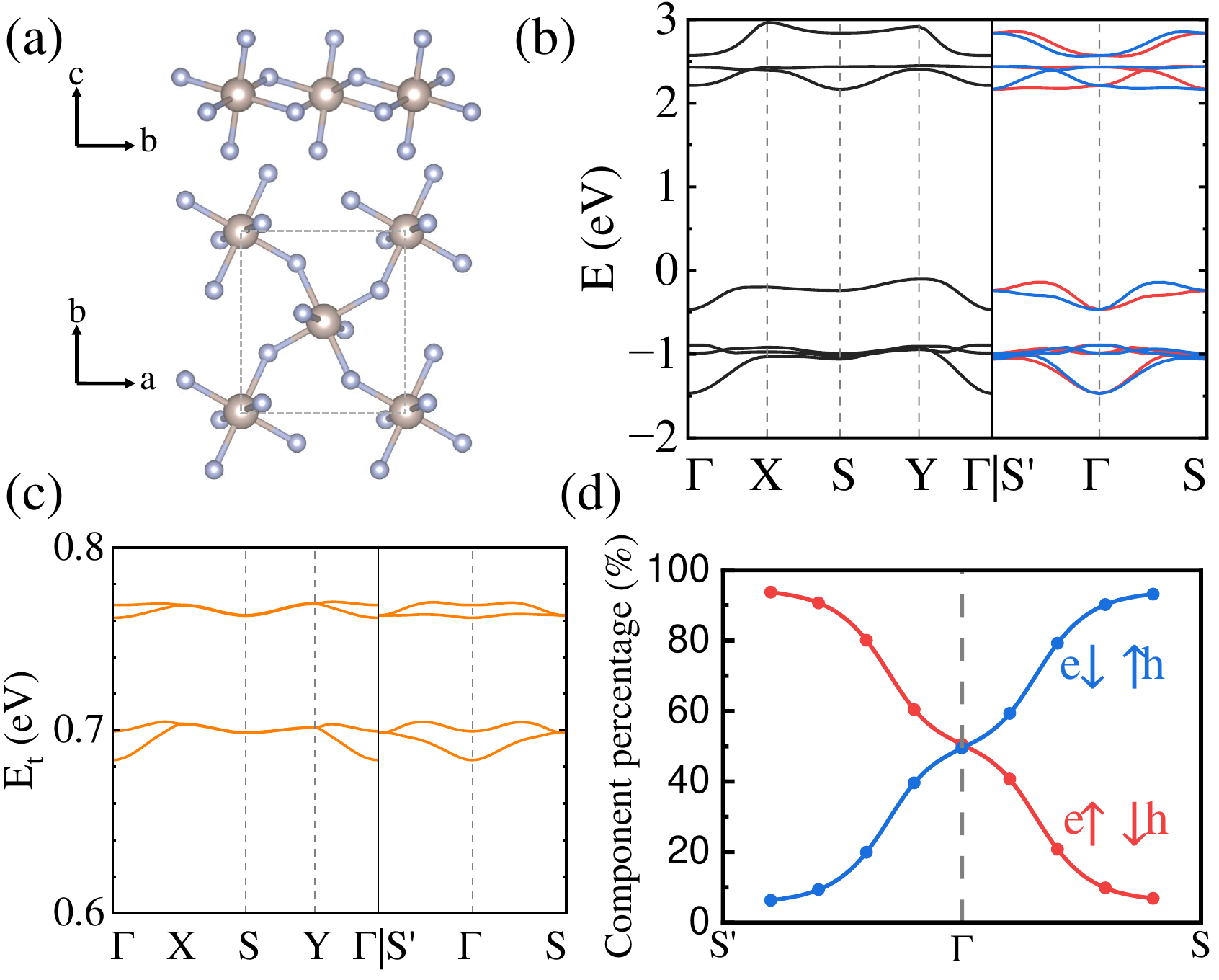}
    \caption{(a) Crystal structure of monolayer $\mathrm{RuF_4}$. (b) Electronic band structure of $\mathrm{RuF_4}$. Spin degeneracy occurs along the $\mathrm{\Gamma-X-S-Y-\Gamma}$ line. The red and blue lines denote the spin-up and spin-down channels, respectively. (c) Excitonic band structure of $\mathrm{RuF_4}$. The high-symmetry path for the center-of-mass momentum $\mathbf{Q}$ is  identical to the $\mathbf{k}$-point path used for electronic structure. (d) The component of the lowest exciton along $\mathrm{S'-\Gamma-S}$. The red line refers to pairs containing spin-up electrons and spin-down holes, while the blue line refers to pairs containing spin-down electrons and spin-up holes. }
    \label{fig:fig2}
\end{figure}

Considering the first block, which describes conventional optical excitons with a spin-conserving configuration $[\psi^*_\uparrow\psi_\uparrow,\psi^*_\downarrow\psi_\downarrow]$, the symmetry-protected degeneracy changes.  When the action of  $U_z(\theta)$ on the basis is considered, the factors from the electron part and the hole part cancel out, and $U_z(\theta)$ acts as the identity matrix. Now the combined $\mathcal{T}U_z(\theta)$ operation gives 
\begin{equation}
    [\mathcal{T}U_z(\theta)]^2\left(\begin{array}{c}
         \psi^*_\uparrow\psi_\uparrow\\
         \psi^*_\downarrow\psi_\downarrow
    \end{array}\right)=\left(\begin{array}{cc}
         \psi^*_\uparrow\psi_\uparrow\\
         \psi^*_\downarrow\psi_\downarrow
    \end{array}\right),
\end{equation}
thus the character $\chi([\mathcal{T}AU_z(\theta)]^2)=2$. Based on the Dimmock–Wheeler character sum rule, we conclude that there is no extra degeneracy. For the second mechanism, with the basis $[\psi^*_\uparrow\psi_\uparrow,\psi^*_\downarrow\psi_\downarrow]$ the $U_z(\phi)$ acts as identity matrix $\mathbf{I_{2\times2}}$, while $U_x(\pi)$ yields an anti-diagonal real matrix
\begin{equation}
    \mathcal{D}(U_z(\phi))=\left(\begin{matrix}1&0\\0&1\end{matrix}\right),
\end{equation}
\begin{equation}
    \mathcal{D}(U_x(\pi))=\left(\begin{matrix}0&1\\1&0\end{matrix}\right).
\end{equation}
In this case, no extra two-dimensional irreducible representation emergies. So neither mechanism protects the extra degeneracy in the excitonic band structure.

Before showing the calculation results, we talk in depth the microscopic nature of the allowed splitting. The single-particle and screened interaction parts only entry the spin diagonal blocks of BSE hamiltonian. The only spin off-diagonal blocks come from the exchange interaction between electron-hole pairs. At the symmetry-invariant momentum, the two diagonal spin sectors of the $M_S=0$ BSE Hamiltonian are related by the collinear spin-space-group symmetry and can therefore be written in the form $H=\left(\begin{smallmatrix}A&X\\X^\dagger&A\end{smallmatrix}\right)$, where $A$ contains the independent-particle contribution together with the spin-conserving direct and exchange terms, while $X$ originates from the exchange-mediated coupling between the two $M_S=0$ spin configurations. The diagonal equivalence of the two spin channels alone would preserve their degeneracy; however, the finite off-diagonal coupling $X$ hybridizes them into symmetry-adapted linear combinations. For $X=X^\dagger$, the Hamiltonian can be exactly transformed into two sectors, $A+X$ and $A-X$. Consequently, the degeneracy is lifted by the exchange coupling.

To examine our symmetry-based theory and find the exotic excitonic phenomena in realistic materials, we perform Green's-function-based many-body perturbation calculations for several examples. The BSE are solved using the Yambo code on top of the density functional theory (DFT) results produced by Quantum Espresso to get the energy levels and band structures of exciton states. The details of the calculation are supplied in the Supplemental Material S1~\cite{SM,kresse_efficiency_1996,giannozzi_advanced_2017,giannozzi_quantum_2009,marini_yambo_2009,sangalli_many-body_2019,hamann_optimized_2013,sangalli_lumen_2026,chen_enumeration_2024,jiang_enumeration_2024,liu_spin-group_2022,xiao_spin_2024,song_constructions_2025}. Due to the reduced screening in two-dimensional (2D) materials, excitonic effects are greatly enhanced, making these systems a promising platform for observing the predicted spliting clearly. Monolayer $\mathrm{RuF_4}$ is a promising candidate as altermagnet. The space group for $\mathrm{RuF_4}$ is $P2_1/c$ (No.14) and the spin space group is classified as $P\prescript{\bar{1}}{}{2_1}/\prescript{\bar{1}}{}{c}\prescript{\infty m}{}{1}$. The crystal structure and electronic band structure is shown in Fig.~\ref{fig:fig2} (a) and (b). The little group at the $\mathrm{\Gamma-X-S-Y-\Gamma}$ line contains the $\{\mathcal{T}||m|\tau\}$ operation, which provides extra spin degeneracy. After solving the BSE at every $\mathbf{Q}$ points, the excitonic band structure is plotted in Fig.~\ref{fig:fig2} (c). The degeneracy along $\mathrm{\Gamma-X}$ and $\mathrm{\Gamma-Y}$ line in electronic band structure is lifted in excitonic band, as predicted by symmetry analysis. One may notice that the degeneracy along $\mathrm{X-S-Y}$ line still exists. We claim that this degeneracy comes from the factor system of the projective irreducible representations at the boundary of the Brillouin zone and is not the result of extra degeneracy caused by $\{\mathcal{T}||A\}$ or $\{U_x(\pi)||A\}$ in altermagnets. The details of the factor system and corresponding degeneracy can be found in the Supplemental Material S2~\cite{SM}. In Fig.~\ref{fig:fig2}(d), the percentage of each component in the lowest exciton at $\mathbf{k}$ points along $\mathrm{S'-\Gamma-S}$ (red for spin-up electron and spin-down hole, while blue for spin-down electron and spin-up hole) is shown, clearly showing the alternating origin of alter-excitons in altermagnets. 

\begin{figure}
    \centering
    \includegraphics[width=1.0\linewidth]{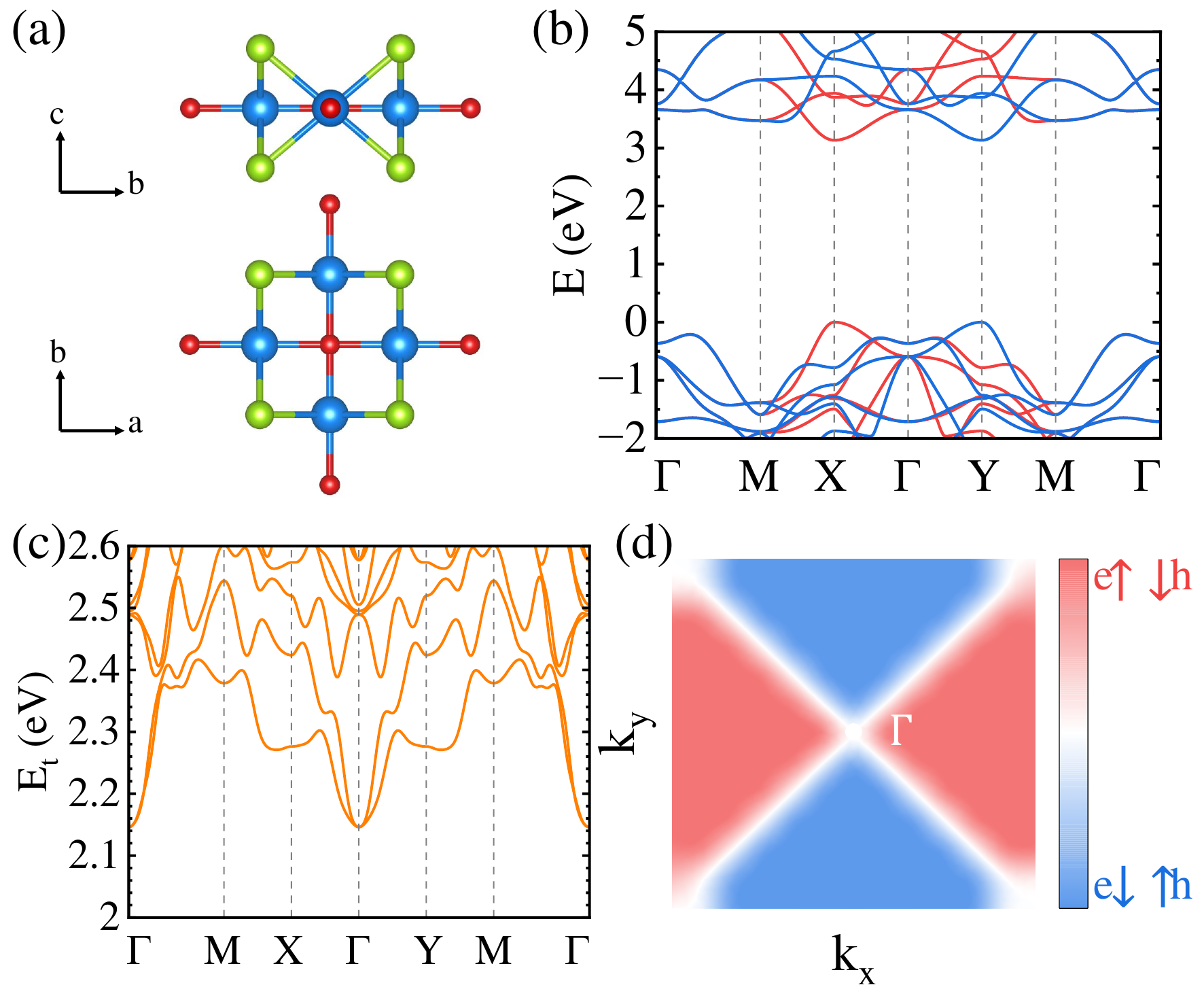}
    \caption{(a) Crystal structure of monolayer $\mathrm{V_2Se_2O}$. (b) Electronic band structure of altermagnets monolayer $\mathrm{V_2Se_2O}$. Spin degeneracy occurs along the $\mathrm{\Gamma-M}$ line. The red and blue lines denote the spin-up and spin-down channels, respectively. (c) Excitonic band structure of monolayer $\mathrm{V_2Se_2O}$. The high-symmetry path for the center-of-mass momentum $\mathbf{Q}$ is $\mathrm{\Gamma-M-X-\Gamma-Y-M-\Gamma}$, identical to the $\mathbf{k}$-point path used for electronic structure. (d) The component of the lowest exciton at each $\mathbf{Q}$ in Brillouin zone. The component distribution also shows d-wave pattern. }
    \label{fig:fig3}
\end{figure}

The monolayer $\mathrm{V_2Se_2O}$ is a direct-gap altermagnet with space group $P4/mmm$ (No.123). The corresponding spin space group is $P\prescript{\bar{1}}{}{4}/\prescript{1}{}{m}\prescript{1}{}{m}\prescript{\bar{1}}{}{m}\prescript{\infty m}{}{1}$. The symmetry operation $\{\mathcal{T}||C_4\}$ connects the opposite-spin sublattices, giving rise to altermagnetism. Previous work has focused on the zero-momentum ($\mathbf{Q}\rightarrow 0$) excitons and their properties. Here, we present the full exciton dispersion along the high-symmetry path in Fig.~\ref{fig:fig3}(c). A clear splitting appears along the $\mathrm{\Gamma-M}$ path, in stark contrast to the strictly protected spin degeneracy in electronic band structure. Plotting the component of the lowest exciton at each $\mathbf{Q}$, where red represent that the exciton consists of a spin-up electron and a spin-down hole and blue the opposite, the spin-momentum locking in exciton dispersion is clearly shown.

\begin{figure*}[!htb]
    \centering
    \includegraphics[width=1.0\linewidth]{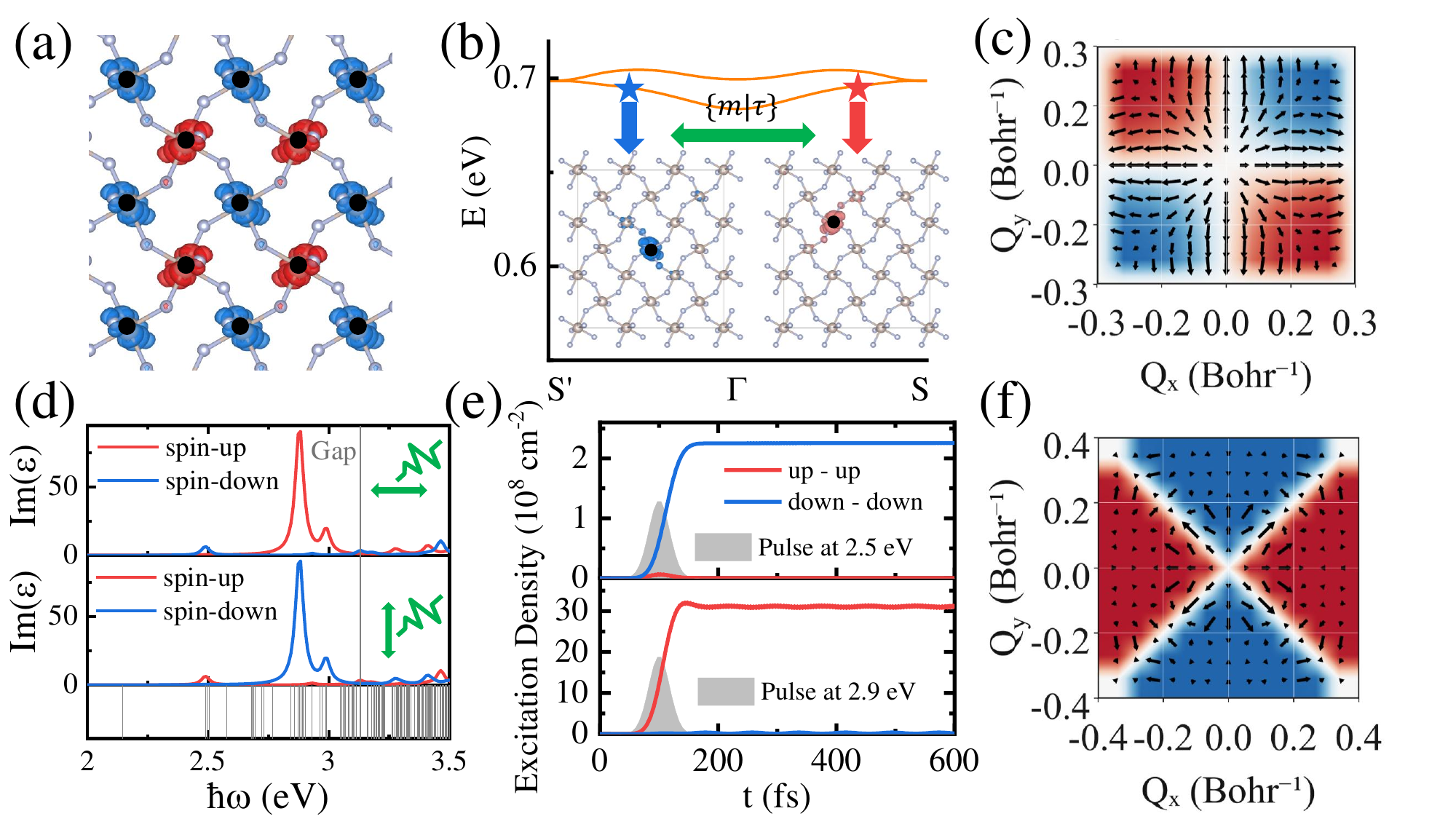}
    \caption{(a) Frenkel-like components of the lowest-energy $\Gamma$-point exciton in monolayer $\mathrm{RuF_4}$. (b) Exciton component at $\{m|\tau\}$-operation-connected $\mathbf{Q}$-points in $\mathrm{RuF_4}$. (c) Group velocity field and alternating nature for 1st exciton band in $\mathrm{RuF_4}$. (d) Spin-resolved absorption spectra of monolayer $\mathrm{V_2Se_2O}$. (e) The spin-resolved excited carrier density at different frequency in $\mathrm{V_2Se_2O}$. (f) Group velocity field and alternating nature for 1st exciton band in $\mathrm{V_2Se_2O}$.}
    \label{fig:fig4}
\end{figure*}

After the analysis in reciprocal space, we show the real-space distribution of alter-excitons. In Fig.~\ref{fig:fig4}(a), we show the Frenkel-like components of the lowest-energy $\Gamma$-point exciton in monolayer $\mathrm{RuF_4}$, which can be clearly visualized. The $e\uparrow h\downarrow$ components are connected to $e\downarrow h\uparrow$ components by real-space $\{m|\tau\}$ operation which exhibits the altermagnetism. For excitons at $\{m|\tau\}$-operation-connected $\mathbf{Q}$-points, for example $Q_1=[0.3,0.3,0]$ and $Q_2=[-0.3,0.3,0]$, the $\{m|\tau\}$ operation would also connect the $e\uparrow h\downarrow$ components of $Q_1$ and the spin-opposite components at $Q_2$, as shown in Fig.~\ref{fig:fig4}(b). 

Finally, using the exciton dispersion and distribution obtained from BSE calculations, we reveal a variety of exotic physical phenomena driven by alter-excitons in altermagnets, highlighting the broad perspectives of this emerging field. The first is the photoexcitation physics. Using the spin-resolved results of BSE and TD-aGW, we clearly show the light polarization-related spin excitation. Especially, in Fig.~\ref{fig:fig4}(d), we show the spin-resolved absorption spectra of monolayer $\mathrm{V_2Se_2O}$. At each excitation frequency, $x$- and $y$-polarized light selectively excite opposite spin channels. Moreover, the exciton envelope wavefunction provides an additional degree of freedom for modulating the absorption spectrum. At the single-particle level, the low-energy optical excitations all lie above the band gap and are determined solely by the symmetry and distribution of the electronic states. In the present case, $x$-polarized light couples predominantly to the spin-up channel, whereas $y$-polarized light mainly couples to the spin-down channel, which is shown in the Supplemental Material Fig.. When considering the excitonic effects, the low-lying excitations are indeed the excitonic states below the band gap. At photon energy 2.7-3.0 eV, the $x$-polarized and $y$-polarized light are coupled to the spin-up channel and spin-down channel, respectively. But at the lowest bright state around 2.5 eV, the $x$-polarized light excites the spin-down channel carriers, whereas the $y$-polarized light excites the spin-up channel, in direct contrast to the single-particle picture. Performing the TD-aGW level real-time propagation using the BSE kernel, we plot the spin-resolved excited carrier density. The pulse is set to be a sine wave centered at 2.5 eV or 2.9 eV with a Gaussian wave packet. The results strongly support our analysis that the photon energy and polarization can simultaneously tune the spin excitations. These distinctive excitonic properties point toward an emerging research direction centered on alter-excitons in altermagnets.

The second is the exciton dynamics coming from the dispersion. Electron-hole pairs with different spin configurations exhibit different group velocities due to spin-momentum locking in altermagnets, resulting in spin-anisotropic exciton evolution. In Fig.~\ref{fig:fig4}(c) and Fig.~\ref{fig:fig4}(f), the group velocity is calculated and shown as a vector field for the lowest band of excitons in the monolayer $\mathrm{RuF_4}$ and monolayer $\mathrm{V_2Se_2O}$, respectively. The background color represents the spin composition of the exciton at each center-of-mass momentum $\mathbf Q$, while the arrows indicate the group velocity of the lowest exciton band. The resulting $d$-wave spin texture gives rise to pronounced spin-selective propagation. For example, in $\mathrm{RuF}_{4}$, a Gaussian exciton wavepacket centered around $\Gamma$ contains finite-$\mathbf Q$ components dominated by different spin configurations: the $e\uparrow h\downarrow$ components propagate preferentially along the $\pm[1,\bar{1}]$ directions, whereas the $e\downarrow h\uparrow$ components propagate along the $\pm[1,1]$ directions. These finite-momentum components separate along the two orthogonal diagonals. An analogous spin-selective propagation occurs in $\mathrm{V}_{2}\mathrm{Se}_{2}\mathrm{O}$. The $e\uparrow h\downarrow$ components propagate preferentially along the $\pm x$ directions, whereas the $e\downarrow h\uparrow$ components propagate along the orthogonal \(\pm y\) directions. Thus, alter-excitons exhibit spin-selective exciton transport, with propagation directions modulated by their distinct momentum-space spin textures.

In summary, by using the recently developed and enriched spin space group theory, we investigate the band structure, momentum-space symmetry, and real-space distribution of alter-excitons in altermagnets. The dispersion and distribution of excitons obtained from the BSE calculation strongly support our symmetry-based analysis. As fundamental neutral excitations in altermagnets, alter-excitons exhibit richer physics than their single-particle counterparts. Our BSE and TD-aGW calculations reveal spin-selective optical absorption that can be controlled by both light polarization and excitation frequency. Furthermore, the spin-selective propagation of excitonic wave packets demonstrates a distinctive transport behavior unique to alter-excitons. Our study establishes alter-excitons as a promising platform for exploring the interplay between magnetism, excitonic many-body effects, and spin-dependent optical and transport phenomena, opening a new avenue for research in altermagnetism.

\bibliography{reference}% Produces the bibliography via BibTeX.

\end{document}